\documentclass[10pt,journal]{IEEEtran}
\def\BibTeX{{\rm B\kern-.05em{\sc i\kern-.025em b}\kern-.08em
    T\kern-.1667em\lower.7ex\hbox{E}\kern-.125emX}}
\usepackage{amsmath}
\usepackage{amsthm}
\usepackage{amsfonts}
\usepackage{amssymb}
\usepackage{graphicx}
\usepackage{cite}
\usepackage{subfigure}
\usepackage{balance}
\usepackage{algorithm}
\usepackage{algorithmic}
\usepackage{enumerate}
\usepackage{color}
\usepackage{array}
\usepackage{indentfirst}
\usepackage{multirow}
\usepackage{booktabs}
\usepackage{color}
\usepackage{slashbox}
\usepackage{diagbox}
\usepackage{threeparttable}
\usepackage[svgnames,table]{xcolor}
\usepackage{bm}
\usepackage{subfigure}
\usepackage{epstopdf}

\usepackage{xcolor}
\usepackage{xpatch}

\makeatletter

\newcommand{\Rmnum}[1]{\expandafter\@slowromancap\romannumeral #1@}

\makeatother

\begin{document}
\title{ Deep Residual Learning-Assisted Channel Estimation in Ambient Backscatter Communications }

\author{\IEEEauthorblockN{Xuemeng Liu, Chang Liu, \emph{Member, IEEE}, Yonghui Li, \emph{Fellow, IEEE}, Branka Vucetic, \emph{Fellow, IEEE}, and \\  Derrick Wing Kwan Ng, \emph{Senior Member, IEEE} } % <-this % stops a space

\thanks{X. Liu, Y. Li, and B. Vucetic are with the School of Electrical and Information Engineering, The University of Sydney, Sydney, NSW 2006, Australia (email: xuemeng.liu.ac@gmail.com, \{yonghui.li, branka.vucetic\}@sydney.edu.au).
}

\thanks{C. Liu and D. W. K. Ng are with the School of Electrical Engineering and Telecommunications, The University of New South Wales, Sydney, NSW 2052, Australia (email: \{chang.liu19, w.k.ng\}@unsw.edu.au).
}
\vspace{-1.0 cm}
}

\maketitle

\begin{abstract}
Channel estimation is a challenging problem for realizing efficient ambient backscatter communication (AmBC) systems. In this letter, channel estimation in AmBC is modeled as a denoising problem and a convolutional neural network-based deep residual learning denoiser (CRLD) is developed to directly recover the channel coefficients from the received noisy pilot signals. To simultaneously exploit the spatial and temporal features of the pilot signals, a novel three-dimension (3D) denoising block is specifically designed to facilitate denoising in CRLD.
In addition, we provide theoretical analysis to characterize the properties of the proposed CRLD.
Simulation results demonstrate that the performance of the proposed method approaches the performance of the optimal minimum mean square error (MMSE) estimator with perfect statistical channel correlation matrix.
\end{abstract}


\vspace{-0.15cm}
\section{Introduction\label{sect: intr}}
%\vspace{-0.10cm}
One of the well-known challenges in unleashing the potential of Internet-of-Things (IoT) is the limitations of energy sources \cite{wong2017key}. In practice, a large number of sensors equipped with limited energy storage create a system performance bottleneck in realizing sustainable communications.
Recently, ambient backscatter communication (AmBC) has been proposed as a promising technique to relieve the energy shortage problem of IoT networks. Specifically, in an AmBC system, a passive tag or sensor could communicate with other nodes by backscattering the ambient RF signals such as television broadcast signals and Wi-Fi signals, instead of directly emitting radio-frequency (RF) signals by itself \cite{van2018ambient}. In fact, by switching to backscattering state or non-backscattering state, a tag or sensor can perform information transmission and utilize the spectral resources of existing systems without requiring additional power.
Thus, AmBC technology has attracted vast attention from academia and industry \cite{hoang2020ambient, liu2020transfer, Liu2020con} in recent years.

%The performance of AmBC systems relies on accurate channel estimation.
%Accurate channel estimation dramatically improves the performance of AmBC systems.
The performance of AmBC systems can be dramatically improved by accurate channel estimation.
%Accurate channel estimation contributes to the performance improvement of AmBC systems.
However, channel estimation problem for AmBC is different from that of traditional communication systems due to the following factors:
\begin{enumerate}[(a)]
  \item Since a passive tag is unable to independently transmit RF signals, the generation of pilot signals for channel estimation requires the cooperation of the RF source;
  \item The channel coefficients of a tag under ON state (backscattering) are not consistent with those under OFF state (non-backscattering).
\end{enumerate}

Therefore, practical channel estimation schemes for AmBC systems need to be redesigned. For example, the optimal minimum mean square error (MMSE) estimator \cite{kay1993fundamentals} cannot be implemented in AmBC systems due to the lack of a precise statistical channel correlation matrix. Thus, a blind expectation maximization (EM)-based method was designed to estimate the absolute values of the channel coefficients \cite{ma2018blind}. However, its estimation performance is unsatisfactory due to the lack of RF source knowledge.
As a remedy, pilots-based methods have been developed. For instance, Ma \emph{et al.} \cite{ma2018machine} designed an EM-aided machine learning scheme. Besides, Zhao \emph{et al.} \cite{zhao2019channel} studied the channel estimation for AmBC with a massive-antenna reader and designed a channel estimation algorithm to jointly estimate the channel coefficients and the directions of arrivals. Although these two methods further improve the estimation performance, there is still a considerable performance gap between them and the optimal MMSE estimator.

Note that the pilot-based channel estimation problem can be considered as a denoising problem \cite{he2018deep, liu2020deep}. Meanwhile, the deep residual learning (DReL) has recently been proposed as a promising denoising technique \cite{he2016deep}.
Motivated by this, different from the existing deep learning based methods adopting deep neural networks to recover channel coefficients, e.g., \cite{huang2020reconfigurable, huang2019deep, chun2019deep, kang2018deep}, we model the channel estimation in AmBC as a denoising problem and develop a DReL approach exploiting a convolutional neural network (CNN)-based deep residual learning denoiser (CRLD) for channel estimation. In CRLD, a three-dimension (3D) denoising block is specifically designed to explore both the spatial and temporal correlations of the received pilot signals. The proposed method inherits the superiorities of CNN and DReL in feature extraction\cite{liu2019JSAC} and denoising to improve the estimation accuracy.
Simulations are conducted and our results show that the proposed method achieves almost the same normalized mean square error (NMSE) performance as the optimal MMSE estimator with the perfect knowledge of the statistical channel correlation matrix.

%\emph{Organization}: Section \Rmnum{2} introduces the system model. In Section \Rmnum{3}, a CRLD is developed to estimate the channel coefficients. Simulation results are presented in Section \Rmnum{4}, and Section \Rmnum{5} finally concludes this letter.

\emph{Notations}: Superscript $T$ represents the transpose. Term ${\mathcal{N}}( \bm{\mu},\mathbf{\Sigma} )$ denotes the Gaussian distribution with a mean vector $\bm{\mu}$ and a covariance matrix $\mathbf{\Sigma}$. Terms ${\bf{I}}_M$ and ${\mathbf{0}}$ represent the $M$-by-$M$ identity matrix and the zero vector, respectively. $\mathbb{R}$ indicates the set of real numbers. $E(\cdot)$ is the statistical expectation. $\|\cdot\|_2$ and $\|\cdot\|_F$ denote the Euclidean norm of a vector and the Frobenius norm of a matrix, respectively. $\max(a,b)$ represents the maximum value between $a$ and $b$.

\begin{figure}[t]
  \centering
  \includegraphics[width=0.56\linewidth]{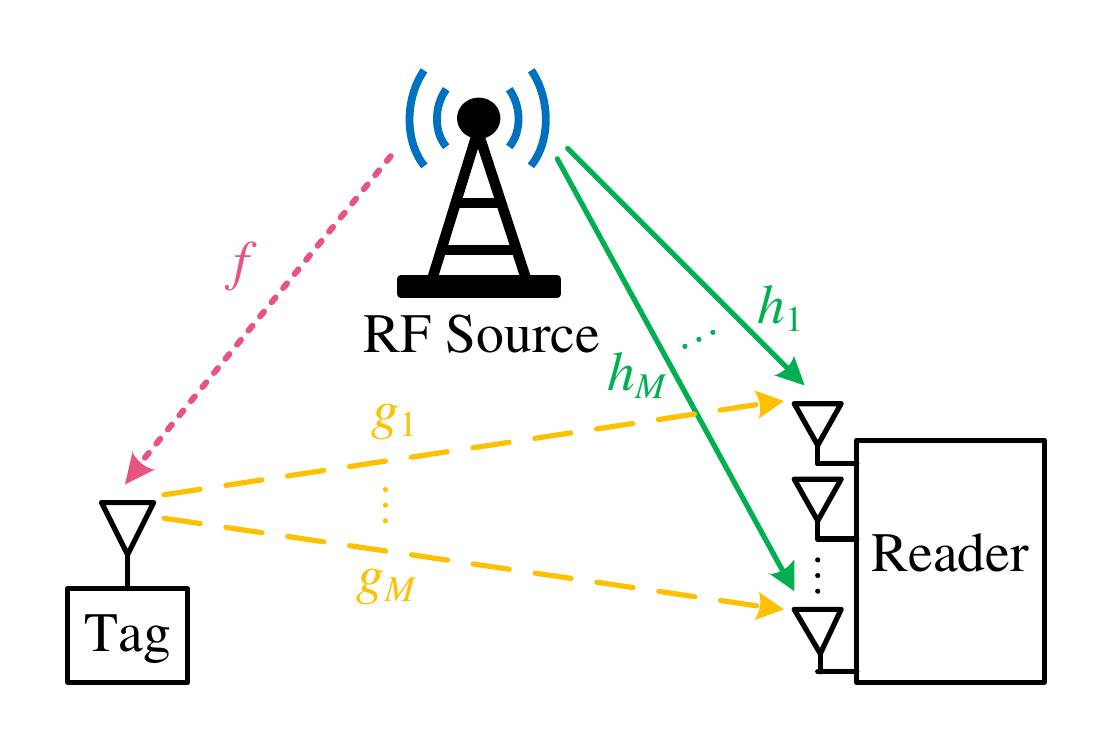}\vspace{-0.1cm}
  \caption{ The considered AmBC system. }\vspace{-0.56cm}\label{Fig:System_model}
\end{figure}

\vspace{-0.1 cm}
\section{System Model}
In this letter, we consider a typical AmBC system, where a single-antenna RF source is surrounded by a single-antenna passive tag and a reader equipped with an $M$-element antenna array, as shown in Fig. \ref{Fig:System_model}.
Although the passive tag is unable to send RF signals by itself, it can transmit its binary tag symbols by deciding whether to reflect the ambient signals (symbol ``1") to the reader or to absorb them (symbol ``0"). Correspondingly, the reader can then recover the binary tag symbols from the received signals. Denote by $\mathbf{y}(n)={{[y_{1}(n),y_{2}(n),\cdots ,y_{M}(n)]}^{T}}, n \in \{ 0,1,\cdots,N_T-1\}$, the $n$-th sampling vector at the reader, where $N_T$ is the number of samples for each frame and $y_m(n)$, $m \in \{1,2, \cdots, M\}$, denotes the received sample from the $m$-th antenna element. The received sampling vector at the reader is expressed as
%\textcolor[rgb]{0.00,0.00,1.00}{\footnotemark}\footnotetext{\textcolor[rgb]{0.00,0.00,1.00}{Different from reconfigurable intelligent surfaces (RIS) systems \cite{huang2019reconfigurable} which adopt a RIS with multiple reflecting elements to assist communication, AmBC systems adopt a low hardware complexity single-antenna tag to transmit its information through whether to backscatter the RF source signal or not.}}
\begin{equation}
\label{y}
\mathbf{y}(n) = \mathbf{h}s(n) + \alpha f\mathbf{g}s(n)c(n) + \mathbf{u}(n).
\end{equation}
Here, $s(n)$ is the RF signal sample and $c(n)\in \mathcal{C} = \{0,1\}$ denotes the tag binary symbol. $\mathbf{h}=[h_1, h_2, \cdots, h_M ]^T$, of which the element $h_m \in \mathbb{R}$ represents the channel coefficient between the RF source and the $m$-th antenna of the reader{\footnotemark}\footnotetext{Note that it is convenient for a neural network to process real-valued data. Thus, a simplified real-valued model is adopted in this letter, which can be easily extended to a complex-valued model via a similar approach as in \cite{kay1993fundamentals}.}. Similarly, $\mathbf{g}=[g_1, g_2, \cdots, g_M]^T$ and $g_m \in \mathbb{R}$ is the channel coefficient between the tag and the $m$-th antenna of the reader. $\alpha \in \mathbb{R}$ is the constant reflection coefficient of the tag. Considering the channel between the RF source and the tag is dominated by a strong line-of-sight (LoS) due to short communication distance, the corresponding channel coefficient $f \in \mathbb{R}$ can be  assumed to be a constant.
In addition, we assume that the noise vector $\mathbf{u}(n) \in \mathbb{R}^{M\times1}$ is an independent and identically distribution (i.i.d.) Gaussian random vector, i.e., $\mathbf{u}(n)\sim \mathcal{N}( \mathbf{0},\sigma _u^2{{\mathbf{I}}_M} )$, where $\sigma_u^2$ is the noise variance.
Based on the system model, the relative coefficient between the reflection link and the direct link can be defined as $ \zeta = E ( ||\alpha f\mathbf{g}||_2^2 ) / E ( ||\mathbf{h}||_2^2 )$, and the instantaneous signal-to-noise ratio (SNR) of the direct link is defined as $\mathrm{SNR} = E(||\mathbf{h}s(n)||_2^2) / E(||\mathbf{u}(n)||_2^2)$.

Note that the received signal can also be written as
\begin{equation}\label{y_v2}
\mathbf{y}(n) = \left\{\begin{matrix}
 \mathbf{w}s(n)+\mathbf{u}(n),& c(n)=1,\\
 \mathbf{h}s(n)+\mathbf{u}(n),& c(n)=0,
\end{matrix}\right.
\end{equation}
where $\mathbf{w} = \mathbf{h} + \alpha f\mathbf{g}$. In this letter, we consider a general slow fading Rayleigh channel model, i.e., $\mathbf{w} \sim \mathcal{N}( \mathbf{0},\mathbf{R}_\mathbf{w} )$ and $\mathbf{h} \sim \mathcal{N}( \mathbf{0},\mathbf{R}_\mathbf{h} )$, where $\mathbf{R}_\mathbf{w} = E(\mathbf{w}\mathbf{w}^{T})$ and $\mathbf{R}_\mathbf{h} = E(\mathbf{h}\mathbf{h}^{T})$ are the statistical channel correlation matrices of $\mathbf{w}$ and $\mathbf{h}$, respectively. In this case, the objective of channel estimation is to estimate the channel coefficient vectors: $\mathbf{w}$ for $c(n)=1$ and $\mathbf{h}$ for $c(n)=0$.

\begin{figure}[t]
  \centering
  \includegraphics[width=0.8\linewidth]{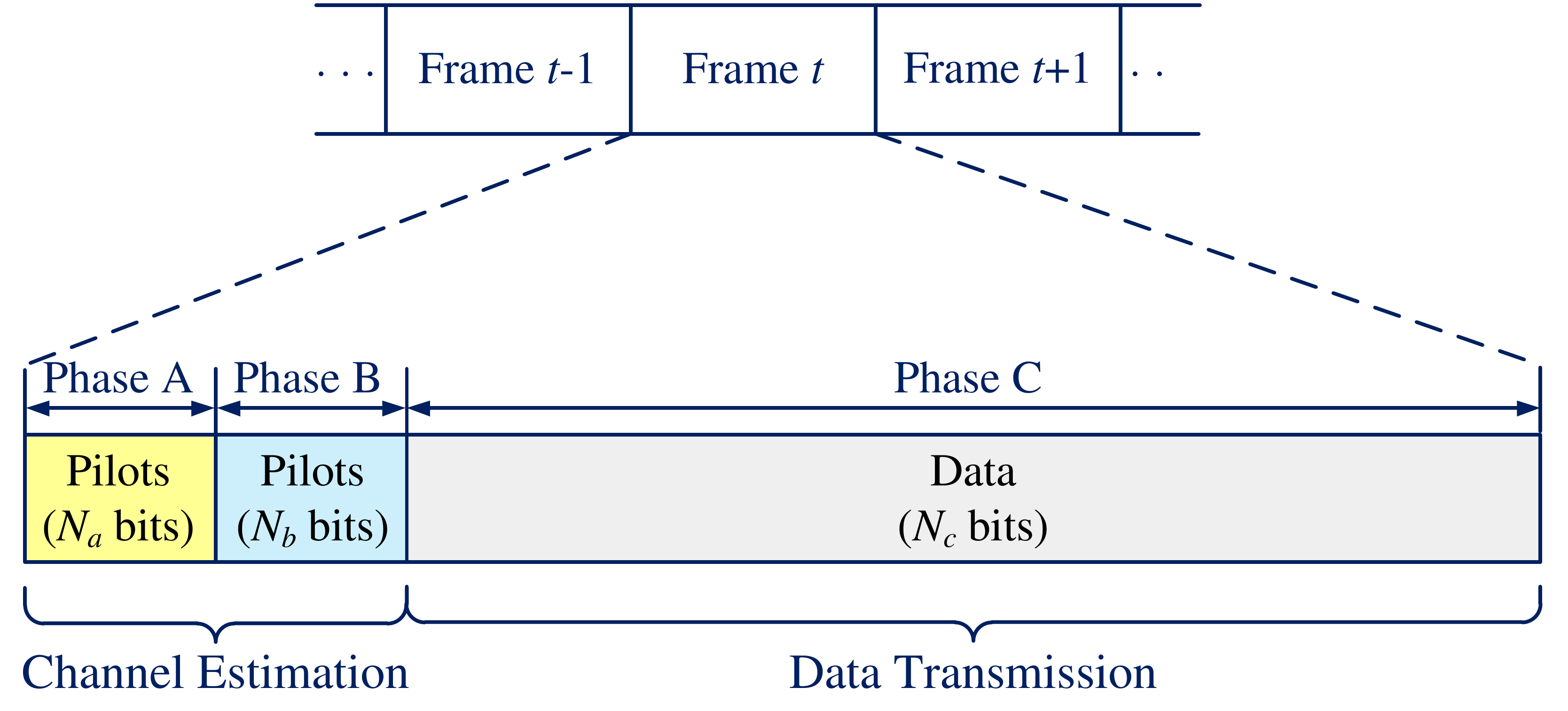}\vspace{-0.1cm}
  \caption{ The designed protocol for the considered AmBC system. }\vspace{-0.68cm}\label{Figure:AmBC-frame}
\end{figure}

Based on this, we design a simple communication protocol for channel estimation in AmBC systems. Assume that there are $T$ frames and each frame has the same structure. As shown in Fig. \ref{Figure:AmBC-frame}, frame $t$, $t \in \{1,2,\cdots,T\}$, consists of three phases: A, B, and C. The first two phases are designed for channel estimation and the remaining phase is for data transmission, which are introduced as follows.
\begin{enumerate}[\title={Phase} A:]
  \item Estimation of $\mathbf{h}$. The tag keeps the state of non-reflection for $N_a$ consecutive sampling periods and the reader estimates $\mathbf{h}$ based on the $N_a$ pilot bits.
  \item Estimation of $\mathbf{w}$. The tag keeps the state of reflection for $N_b$ consecutive sampling periods and the reader estimates $\mathbf{w}$ based on the $N_b$ pilot bits.
  \item Data transmission. The tag transmits $N_c$ information bits by reflecting or absorbing the RF signal.
\end{enumerate}
Note that in the designed protocol, phase A and phase B are adopted to generate pilot bits as the input of the well-trained CRLD to estimate the channel coefficients.
After that, the reader can then decode the received tag symbols in phase C exploiting the estimated channel coefficients.
In the designed protocol, we set $N_a \ll N_c$ and $N_b \ll N_c$ such that there are still enough information bits for data transmission.
Specifically, we set $s(n)=1$ for all the pilot signals and thus the channel estimation becomes a denoising problem, i.e., recovering $\mathbf{x}$ from a noisy observation
\vspace{-0.2cm}
\begin{equation}
\label{y_denoise}
  \mathbf{y}(n) = \mathbf{x} + \mathbf{u}(n), \vspace{-0.2cm}
\end{equation}
where $\mathbf{x} = \mathbf{w}$ or $\mathbf{h}$ depending on $c(n)=1$ or $0$.

In the following, we will develop a denoising algorithm to recover channel coefficients from the received noisy pilot bits.

\begin{figure*}[t]
  \centering
  \includegraphics[width=0.99\linewidth]{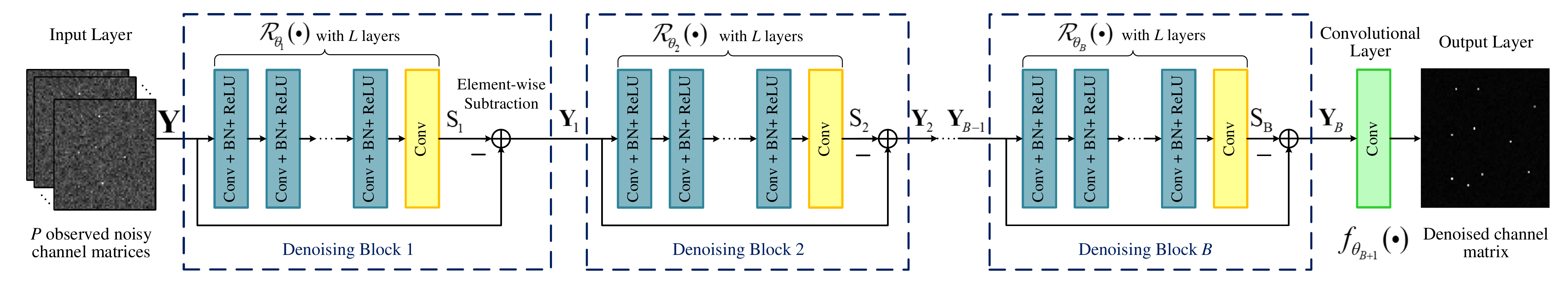}\vspace{-0.2cm}
  \caption{ The proposed CRLD architecture for channel estimation in AmBC. }\vspace{-0.4cm}\label{Figure:CRLD}
\end{figure*}

\vspace{-0.1cm}
\section{CNN-based Residual Learning Denoiser}
In this section, we develop a DReL approach to learn the residual noise to recover the channel coefficients from the noisy pilot signals. Specifically, we adopt CNN to facilitate DReL via proposing a CNN-based deep residual learning denoiser (CRLD). Instead of directly learning a mapping from a noisy channel matrix to a denoised channel matrix, we specifically design a 3D denoising block to learn the residual noise from the noisy channel matrices temporally and spatially for denoising.
In the following, we will introduce the proposed CRLD architecture and the related algorithm, respectively.

\vspace{-0.2cm}
\subsection{CRLD Architecture}
As shown in Fig. \ref{Figure:CRLD}, the CRLD consists of an input layer, $B$ denoising blocks, one convolutional layer, and one output layer. The hyperparameters are summarized in Table I and each layer is introduced as follows.

\begin{table}[t]
\normalsize
%\color{blue}
\caption{Hyperparameters of the proposed CRLD}
\vspace{-0.2cm}
\centering
\small
\renewcommand{\arraystretch}{1.25}
\begin{tabular}{c c c}
  \hline
   \multicolumn{3}{l}{\textbf{Input}: 3D matrix with the size of $M_a \times M_b \times P$} \\
  \hline
   \multicolumn{3}{l}{\textbf{Denoising Block} (CRLD has $B$ identical denoising blocks):} \\
  \hspace{0.4cm} \textbf{Layers} & \textbf{Operations} &  \hspace{0.4cm} \textbf{Filter Size}   \\
  \hspace{0.4cm} 1 & Conv + BN + ReLU  & \hspace{0.4cm}  $ 64 \times ( 3 \times 3 \times P ) $   \\
  \hspace{0.4cm} $ 2 \sim L-1 $ & Conv + BN + ReLU & \hspace{0.4cm}  $ 64 \times ( 3 \times 3 \times 64 ) $   \\
  \hspace{0.4cm} $L$ & Conv & \hspace{0.4cm} $ P \times ( 3 \times 3 \times 64 ) $  \\
  \multicolumn{3}{l}{\textbf{Convolution Layer}: Conv with filter size of $1 \times (M_a \times M_b \times P)$ } \\
  \hline
   \multicolumn{3}{l}{\textbf{Output}: Denoised channel matrix with the size of $M_a \times M_b$} \\
  \hline
\end{tabular}\vspace{-0.6cm}
\end{table}

(a) \textbf{Input Layer}: Assume that there are $P$ observed pilot bits $\{\mathbf{y}(0), \mathbf{y}(1), \cdots, \mathbf{y}(P-1)\}$, $P=N_a$ or $N_b$ as defined in Fig. \ref{Figure:AmBC-frame}, to further explore the spatial and temporal correlations, we first reshape each $\mathbf{y}(p), p \in \{0,1,\cdots,P-1\}$, into a spatial two-dimension (2D) matrix form, denoted by $\mathbf{Y}(p) \in \mathbb{R}^{M_a \times M_b}$, where $1\leq M_a,M_b \leq M$, and $M_aM_b=M$, and then stack them into a 3D matrix as the network input:
\begin{equation}\label{Y}
\mathbf{Y} = \mathcal{F} ( [\mathbf{Y}(0), \mathbf{Y}(1), \cdots, \mathbf{Y}(P-1)] ),
\end{equation}
where $\mathbf{Y} \in \mathbb{R}^{ M_a \times M_b \times P}$ denotes the input of the network and $\mathcal{F}(\cdot)\mathrm{:}~\mathbb{R}^{M_a \times M_bP}\mapsto\mathbb{R}^{M_a \times M_b \times P}$ is the mapping function for stacking matrices.
Note that the proposed CRLD is a universal network structure.
Considering the practical requirement of the coverage area and the adopted carrier frequency of the reader \cite{hoang2020ambient},
we provide a realization of the proposed CRLD where the network input size is set as $M=64$, $M_a=8$, $M_b=8$, and $P=2$.

(b) \textbf{Denoising blocks}: The CRLD has $B$ denoising blocks and each of them has an identical structure, which consists of two types of layers denoted by two different colors, as shown in Fig. \ref{Figure:CRLD}:
(i) Conv+BN+ReLU
%\textcolor[rgb]{0.00,0.00,1.00}{\footnotemark}\footnotetext{\textcolor[rgb]{0.00,0.00,1.00}{Note that the ``Conv + BN + ReLU'' layer is a sequential combination of Conv, BN, and ReLU. When an input $\mathbf{Y}_i$ is sent to the ``Conv + BN + ReLU'' layer, it experiences the operations of Conv, BN, and ReLU sequentially.}}
for the first $1\sim L-1$ layers: convolution\footnotemark\footnotetext{Note that each convolution output is the result of the filter and all the $P$ temporal observations. Thus, the learned features by CRLD contain the temporal information, which contributes to accurate channel estimations.} (Conv) is first operated and then a batch normalization \cite{ioffe2015batch} (BN) is applied after the Conv to facilitate the network training speed and stability. Finally, to enhance the network presentation ability, ReLU is adopted as the activation function which is defined as $y = \max(0,x)$;
(ii) Conv for the last layer: the Conv is adopted to obtain the residual noise $\mathbf{S}_i$ for the subsequent element-wise subtraction.

Therefore, the $i$-th, $i=1,2,\cdots,B$, $L$-layer subnetwork can be modeled as a non-linear function $\mathcal{R}_{\theta_i}(\cdot)$ with parameter $\theta_i$, for each denoising block, we have
\vspace{-0.1cm}
\begin{equation}\label{Y_i}
\mathbf{Y}_{i} = \mathbf{Y}_{i-1} - \mathbf{S}_i = \mathbf{Y}_{i-1} - \mathcal{R}_{\theta_i}(\mathbf{Y}_{i-1}), \forall i. \vspace{-0.1cm}
\end{equation}
Here, $\mathbf{Y}_{0}=\mathbf{Y}$ and $\mathbf{Y}_{i-1}$ and $\mathbf{Y}_{i}$ denote the input and output of the $i$-th denoising block, respectively. In addition, $\mathbf{S}_i = \mathcal{R}_{\theta_i}(\mathbf{Y}_{i-1})$ is the residual term between $\mathbf{Y}_{i-1}$ and $\mathbf{Y}_{i}$, and thus it is called as the residual noise in the literature.

(c) \textbf{Convolutional layer}: A convolutional layer, denoted by the green color box in Fig. \ref{Figure:CRLD}, is added between the last denoising block and the network output to combine the $P$ denoised channel matrices to reconstruct an $M_a$-by-$M_b$ output.

In summary, to further improve the denoising performance for channel recovery, the proposed CRLD architecture adopts $B$ denoising blocks to remove the noise gradually and finally exploits a Conv layer to reconstruct the output.

\vspace{-0.1cm}
\subsection{CRLD-based Estimation Algorithm}
Based on the proposed CRLD architecture, we then design a CRLD-based channel estimation scheme, which consists of offline training phase and online estimation phase.

\subsubsection{Offline Training Phase}
Given a training set
\begin{equation}\label{}
  (\Omega_\mathbf{Y}, \Omega_\mathbf{X}) \hspace{-0.1cm}=\hspace{-0.1cm} \big\{\hspace{-0.05cm}(\mathbf{Y}^{(1)},\mathbf{X}^{(1)}),\cdots,(\mathbf{Y}^{(K)},
  \mathbf{X}^{(K)})\hspace{-0.05cm}\big\}.
\end{equation}
Here, $(\mathbf{Y}^{(k)},\mathbf{X}^{(k)})$, $k \in \{ 1,2,\cdots,K\}$, denotes the $k$-th example of the training set. $\mathbf{Y}^{(k)} \in \mathbb{R}^{ M_a \times M_b \times P}$ is the network input, as defined in (\ref{Y}). $\mathbf{X}^{(k)} \in \mathbb{R}^{ M_a \times M_b }$ is the label which is the matrix form of $\mathbf{w}$ or $\mathbf{h}$, as defined in (\ref{y_denoise}).

According to the MMSE criterion \cite{kay1993fundamentals}, the cost function of the offline training phase can be expressed as \vspace{-0.2cm}
\begin{equation}\label{}
  J_{\mathrm{MSE}}(\theta) =  \sum\limits_{k=1}^{K}||\mathbf{X}^{(k)} - \tilde{\mathbf{X}}^{(k)}_{\theta}||_F^2, \vspace{-0.2cm}
\end{equation}
where $\tilde{\mathbf{X}}^{(k)}_{\theta}$ is the output of CRLD.
We can then use the backpropagation (BP) algorithm \cite{goodfellow2016deep} %{\color{blue}\footnotemark\footnotetext{\color{blue}Note that the backpropagation algorithm is universally applicable for different network architectures \cite{goodfellow2016deep}. Therefore, applying the backpropagation to the CRLD architecture is the same as that on a general network, i.e., updating the CRLD network weights from the last layer to the first layer progressively.}}
to progressively update the network weights and finally obtain a well-trained CRLD:
\begin{equation}\label{CRLD}
  h_{\theta^*}(\mathbf{Y}) = f_{\theta^*_{B+1}}\left(\mathbf{Y} - \sum\limits_{i = 1}^{B}\mathcal{R}_{\theta_i^*}\left(\mathbf{Y}_{i-1}\right)\right),
\end{equation}
where $h_{\theta^*}(\cdot)$ denotes the expression of the well-trained CRLD with the well-trained weight $\theta^*=\{\theta_1^*,\theta_2^*,\cdots,\theta_{B+1}^*\}$.

\subsubsection{Online Estimation Phase}
Given the pilot-based test data $\mathbf{Y}_{\mathrm{test}}$, we can then directly obtain the output of CRLD: $\tilde{\mathbf{X}}_{\mathrm{CRLD}} = h_{\theta^*}(\mathbf{Y}_{\mathrm{test}})$. Reshaping $\tilde{\mathbf{X}}_{\mathrm{CRLD}}$ into a $M$-by-$1$ vector $\tilde{\mathbf{x}}_{\mathrm{CRLD}}$, we finally obtain the estimated channel coefficient vector, denoted by $\tilde{\mathbf{w}} = \tilde{\mathbf{x}}_{\mathrm{CRLD}}$ or $\tilde{\mathbf{h}}=\tilde{\mathbf{x}}_{\mathrm{CRLD}}$.

\subsubsection{Algorithm Steps}
The proposed algorithm is summarized in \textbf{Algorithm 1}, where $i$ and $I$ are the iteration index and the maximum iteration number, respectively.
In addition, $I$ is the maximum iteration number which is controlled by an early stopping rule \cite{goodfellow2016deep}.

\begin{table}[t]
\small
\centering
\begin{tabular}{l}
\toprule[1.8pt] \vspace{-0.4cm}\\
\hspace{-0.1cm} \textbf{Algorithm 1} {CRLD-based Estimation Algorithm}  \\
\toprule[1.8pt] \vspace{-0.3cm}\\
\textbf{Initialization:} $i = 0$ \\
\textbf{Offline Training Phase:} \\
1:\hspace{0.75cm}\textbf{Input:} Training set $({\Omega }_{\mathbf{Y}},{\Omega }_{\mathbf{X}})$\\
2:\hspace{1.1cm}\textbf{while} $i \leq I $ \textbf{do} weights update: \\
3:\hspace{1.6cm}Update $\theta$ by BP algorithm on $J_{\mathrm{MSE}}(\theta)$ \\
\hspace{1.8cm} $i = i + 1$  \\
4:\hspace{1.1cm}\textbf{end while} \\
5:\hspace{0.75cm}\textbf{Output}: Well-trained CRLD ${h}_{\theta^*}( \cdot )$ as defined in (\ref{CRLD})\\
\textbf{Online Estimation Phase:} \\
6:\hspace{0.6cm}\textbf{Input:} Test data $\mathbf{Y}_{\mathrm{test}}$ \\
7:\hspace{0.95cm}\textbf{do} channel estimation with CRLD ${h}_{\theta^*}( \cdot )$  \\
8:\hspace{0.6cm}\textbf{Output:} $\tilde{\mathbf{X}}_{\mathrm{CRLD}}$, i.e., $\tilde{\mathbf{w}}$ or $\tilde{\mathbf{h}}$. \\
\bottomrule[1.8pt]
\end{tabular}\vspace{-0.5cm}
\end{table}

\vspace{-0.4cm}
\subsection{Theoretical Analysis}
To offer more insight of the proposed CRLD method, we then analyze the output of CRLD and characterize its properties theoretically.
Note that the BN and the ReLU operations are adopted for enhancing the training speed and the network stability. Thus, we mainly investigate the effect of the convolution operations on the CRLD output.
For the convenience of analysis, let $\tilde{M}_b = PM_b$, a 2D matrix \vspace{-0.1cm}
\begin{equation}\label{Y_2D}
  \tilde{\mathbf{Y}} = [\mathbf{Y}(0), \mathbf{Y}(1), \cdots, \mathbf{Y}(P-1)]  \in \mathbb{R}^{ M_a \times {\tilde{M}_b}} \vspace{-0.1cm}
\end{equation}
is considered as the input of CRLD and it can be easily extended to the 3D input based on (\ref{Y}).
%Let $\tilde{\mathbf{Y}}_{i} \in \mathbb{R}^{ M_a \times {\tilde{M}_b}}$ denote the 2D matrix form of $\mathbf{Y}_{i}$, $i = 0,1, \cdots, 8$,
Since the convolution operation can be formulated as a production of two matrices \cite{goodfellow2016deep}, the well-trained subnetwork can be expressed as
\begin{equation}\label{R_analysis_v1}
  \mathcal{R}_{\theta_i}(\tilde{\mathbf{Y}}_{i-1}) = \tilde{\mathbf{Y}}_{i-1}\mathbf{W}^*_i,
\end{equation}
where $\mathbf{W}^*_i$ represents the well-trained network weights of the subnetwork.
Thus, based on (\ref{Y_i}), we have $\tilde{\mathbf{Y}}_i = \prod_{j=1}^{i}\tilde{\mathbf{Y}}\tilde{\mathbf{W}}^*_{j},$
where $\tilde{\mathbf{Y}}_0 =\tilde{\mathbf{Y}}$ and $\tilde{\mathbf{W}}^*_j = \mathbf{I}_P - \mathbf{W}^*_i$. In this case, we can rewrite (\ref{R_analysis_v1}) as \vspace{-0.3cm}
\begin{equation}\label{R_analysis_v2}
  \mathcal{R}_{\theta_i}(\tilde{\mathbf{Y}}_{i-1}) = \prod_{j=1}^{i-1}\tilde{\mathbf{Y}}\tilde{\mathbf{W}}^*_j\mathbf{W}^*_i,  \forall i \in \{2,3,\cdots, 8\} \vspace{-0.3cm}
\end{equation}
Thus, $\mathbf{Y}_B$ can be written as \vspace{-0.1cm}
\begin{equation}\label{Y_B_P1}
   \tilde{\mathbf{Y}}_B = \tilde{\mathbf{Y}} - \sum\limits_{i = 1}^{B}\mathcal{R}_{\theta_i}\left(\tilde{\mathbf{Y}}_{i-1}\right) = \tilde{\mathbf{Y}} (\mathbf{I}_{\tilde{M}_b}-\mathbf{W}), \vspace{-0.1cm}
\end{equation}
where $\mathbf{W} = \mathbf{W}_1 + \sum_{i=2}^{B}\prod_{j=1}^{i-1}\tilde{\mathbf{W}}_j\mathbf{W}_i$, and $\mathbf{I}_{\tilde{M}_b}$ denotes the ${\tilde{M}_b}$-by-${\tilde{M}_b}$ identity matrix.
Finally, the well-trained CRLD, i.e., the CRLD-based estimator can be expressed as
\begin{equation}\label{X_CRLD}
  \mathbf{X}_{\mathrm{CRLD}} = \tilde{\mathbf{Y}} (\mathbf{I}_{\tilde{M}_b}-\mathbf{W}^*) \mathbf{W}^*_{B+1},
\end{equation}
where $\mathbf{W}^*_{B+1}$ denotes the well-trained weights of the convolutional layer $f_{\theta_{B+1}}(\cdot)$.
On the other hand, the expression of the optimal MMSE estimator is
\begin{equation}\label{X_MMSE}
  \mathbf{X}_{\mathrm{MMSE}} = \tilde{\mathbf{Y}} \left(\mathbf{I}_{{\tilde{M}_b}} - \alpha \mathbf{S}^H \left(\alpha \mathbf{S} \mathbf{S}^H + \mathbf{R_X}^{-1}\right)^{-1} \mathbf{S}\right)  \alpha \mathbf{S}^H \mathbf{R}_{\mathbf{X}},
\end{equation}
where $\mathbf{X}\in \mathbb{R}^{M_a \times M_b}$ is the matrix form of the channel vector $\mathbf{x}$, $\mathbf{S} = [\mathbf{I}_{M_b},\mathbf{I}_{M_b},\cdots,\mathbf{I}_{M_b}] \in \mathbb{R}^{M_b \times {\tilde{M}_b}}$, $\mathbf{R_X}= E(\mathbf{X}^H \mathbf{X})$ denotes the statistical correlation matrix and $\alpha = \frac{1}{M_a\sigma_u^2}$.
It can been seen from (\ref{X_CRLD}) that the weight matrices $\mathbf{W}^*$ and $ \mathbf{W}_{B+1}^*$ can be learned from the training set through the offline training of the proposed CRLD.
Since we adopt the MMSE-based cost function for the offline training, thus, if the training set is sufficiently large, the proposed CRLD-based estimator can learn and mimic the expression of the optimal estimator under the MMSE criterion \cite{kay1998fundamentals}, i.e., the expression of the optimal MMSE estimator in (\ref{X_MMSE}).
In fact, the proposed CRLD achieves the optimal MMSE performance when the weights approach $\mathbf{W}^*=\frac{1}{M_a\sigma_u^2} \mathbf{S}^H \left(\frac{1}{M_a\sigma_u^2} \mathbf{S} \mathbf{S}^H + \mathbf{R_X}^{-1}\right)^{-1} \mathbf{S}$ and $\mathbf{W}^*_{B+1}=\frac{1}{M_a\sigma_u^2} \mathbf{S}^H \mathbf{R}_{\mathbf{X}}$ for a large enough training set. This will be verified through the simulation results in Section IV.

%\vspace{0.2cm}
\section{Simulation Results}
In this section, we provide simulation results to verify the efficiency of the proposed algorithm. As shown in Fig. 1, a classical AmBC system with a $64$-element multi-antenna reader is considered for simulation. In the simulation, the ambient source is modeled by a Gaussian random variable and Rayleigh channel model is adopted, as defined in (\ref{y_v2}). The hyperparameters of the proposed CRLD are summarized in Table I, where we set $M_a = M_b = 8$, $B = 3$, $L = 8$, and $P = N_a = N_b \in [2,16]$. To evaluate the channel estimation performance, we compare the proposed CRLD method with the optimal MMSE method and the least square (LS) method \cite{kay1993fundamentals}. The normalized MSE (NMSE) is adopted as the performance metric which is defined as \vspace{-0.1cm}
\begin{equation}\label{}
  \mathrm{NMSE} = {E \left( \|\mathbf{x} - \tilde{\mathbf{x}}\|_2^2 \right)} / {E \left( \|\mathbf{x}\|_2^2 \right)}, \vspace{-0.1cm}
\end{equation}
where $\mathbf{x}$ and $\tilde{\mathbf{x}}$ are the ground truth and the estimated value, respectively.
All the presented simulation results are obtained through averaging $100,000$ Monte Carlo realizations.

\begin{figure}[t]
  \centering\vspace{-0.1cm}
  \includegraphics[width=3.1in,height=2.2in]{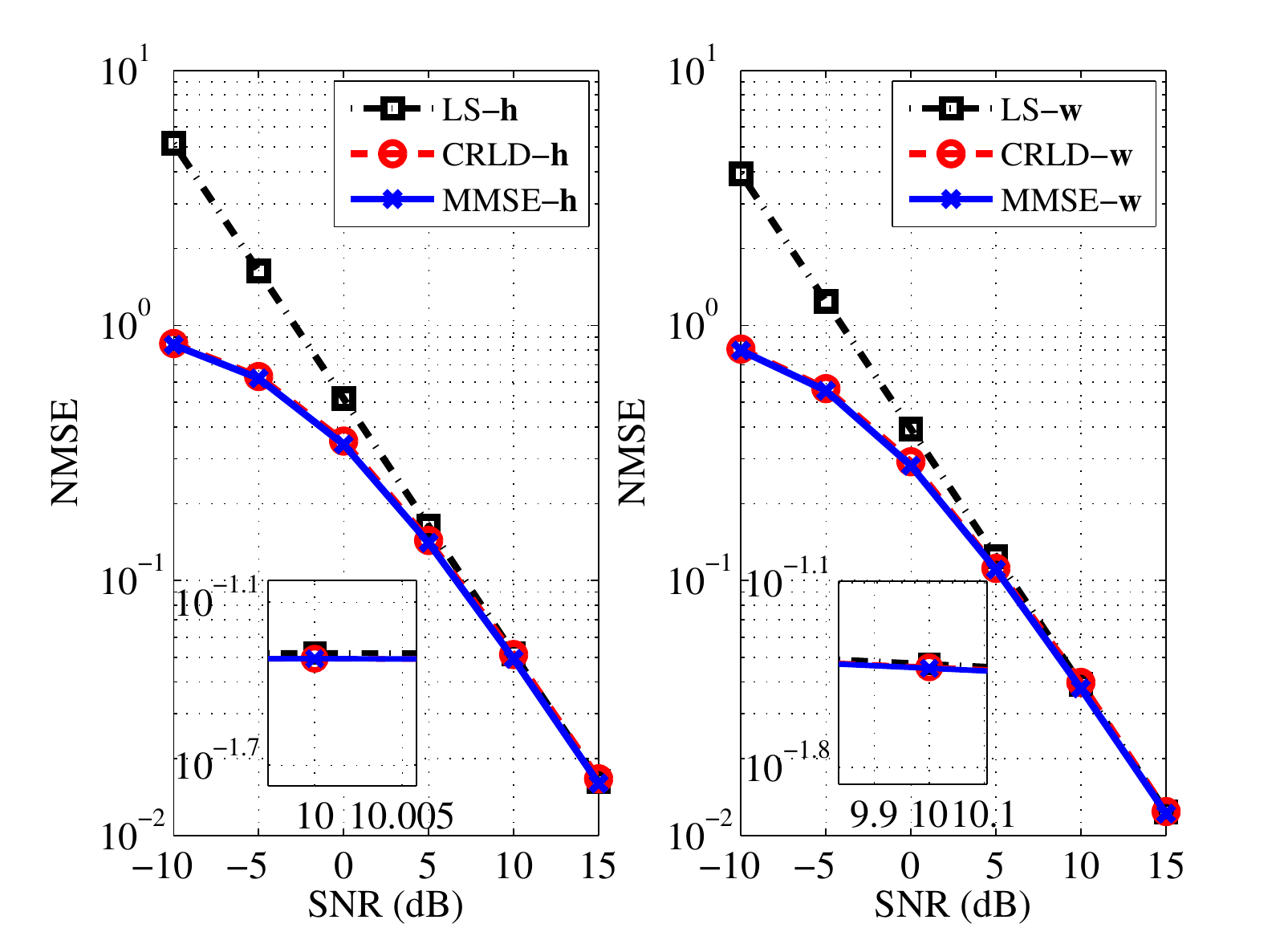}\vspace{-0.1cm}
  \begin{tabular}{l l}
\hspace{0.8cm}{\scriptsize{(a) $\mathbf{h}$ for $c(n)=0$. }} \hspace{1.7cm}{\scriptsize{(b) $\mathbf{w}$ for $c(n)=1$. }} \hspace{0.3cm}\\
\end{tabular}\vspace{-0.26cm}
  \caption{NMSE versus SNR under $\zeta$ = $-5$ dB and $N_a = N_b = 2$.}
  \vspace{-0.2cm}
  \label{Figure:NMSE_SNR}
\end{figure}

We first evaluate the NMSE performance with different SNRs in Fig. \ref{Figure:NMSE_SNR}. Note that the optimal MMSE method requires the perfect statistical channel correlation matrices, which is not always available in practice. Thus, we merely present it for benchmarking and its expression was defined in (\ref{X_MMSE}).
In particular, the proposed CRLD method approaches the optimal MMSE method based on the perfect statistical channel correlation matrix in all considered scenarios.
On the other hand, it can be seen from Fig.~\ref{Figure:NMSE_SNR} that in the high SNR regime, the performance of the LS method approaches that of the optimal MMSE method as the impact of noise is limited. However, the LS method still has a large performance gap compared with the MMSE method and the proposed CRLD method in the low SNR regime. For example, the CRLD can achieve a SNR gain of $4$ $\mathrm{dB}$ in terms of $\mathrm{NMSE} \approx 10^{-0.5}$ compared with the LS method. This is because the LS method treats the channel coefficients as deterministic but unknown constants, while the proposed method and the MMSE method handle the impact of the channel as a random variable. Thus, the latter two schemes can exploit the prior statistical knowledge of the channel matrices to further improve the estimation accuracy.

\begin{figure}[t]
  \centering\vspace{-0.1cm}
  \includegraphics[width=3.1in,height=2.2in]{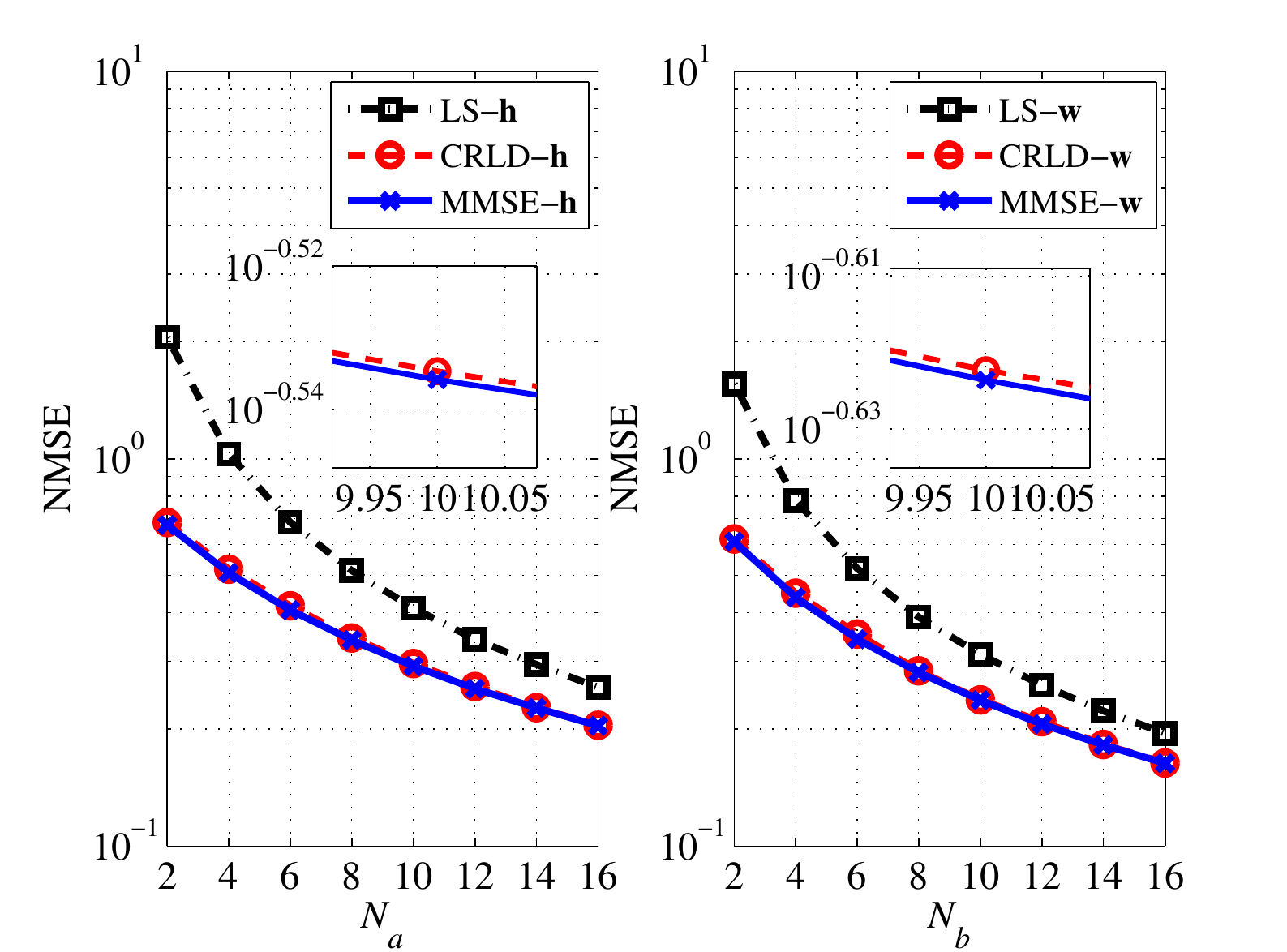}\vspace{-0.1cm}
  \begin{tabular}{l l}
\hspace{0.8cm}{\scriptsize{(a) $\mathbf{h}$ for $c(n)=0$. }} \hspace{1.7cm}{\scriptsize{(b) $\mathbf{w}$ for $c(n)=1$. }} \hspace{0.3cm}\\
\end{tabular}\vspace{-0.26cm}
  \caption{\hspace{-0.1cm}NMSE versus the number of pilots under SNR = $-6$ dB,  $\zeta$ = $-5$ dB.}
  %\vspace{-0.05cm}
  \label{Figure:NMSE_pilot}
\end{figure}

Fig. \ref{Figure:NMSE_pilot} presents the results of NMSE with different numbers of pilots in a noisy communication environment. It is shown that the NMSEs of all the methods decrease with the increasing number of pilot symbols, and the CRLD method can always achieve the same performance as that of the optimal MMSE method. The reason is that our proposed method can efficiently exploit the temporal correlations of the pilot signals for improving the accuracy of channel estimation.

\begin{table}[t]
\normalsize
%\color{blue}
\caption{Computational complexity of different estimation algorithms }
\vspace{-0.1 cm}
\centering
\small
\renewcommand{\arraystretch}{1.25}
\begin{tabular}{c c c}
  \hline
  {\textbf{\footnotesize{Algorithm}}} & {\textbf{\footnotesize{Online Estimation}}} & {\textbf{\footnotesize{Offline Training}}} \\
  \hline
  {\textbf{\footnotesize{LS}}}
  & \footnotesize{$O( MP)$}
  & -
  \\
  {\textbf{\footnotesize{MMSE}}}
  & \footnotesize{$O(P^3+MP^2)$}
  & -
  \\
  {{\textbf{\footnotesize{CRLD}}}}  &
  \scriptsize{{$O\left( BM\sum\limits_{l = 1}^{L}n_{l-1}  s_l^2  n_l\right)$}} &
  \scriptsize{$O\left( N_tIBM\sum\limits_{l = 1}^{L}n_{l-1}  s_l^2  n_l \right)$}
  \\
  \hline
\end{tabular}
\vspace{-0.3cm}
\end{table}

Finally, we investigated the computational complexities of different algorithms and summarize them in Table II. Here, $s_l$ denotes the side length of the $l$-th convolutional layer's filter and $n_{l-1}$ and $n_l$ represent the depthes of the input feature map and the output feature map of the $l$-th convolutional layer, respectively.
In addition, $N_t$ denotes the number of training examples.
It is shown that the computational complexity of the LS and MMSE methods only come from the online detection, while the CRLD has an additional complexity due to the offline training.
Specifically, the LS and MMSE methods have fixed complexities, while the complexity of the CRLD changes with the network size.
Correspondingly, we then execute these algorithms on a PC with a i7-8700 3.20 GHz CPU and a Nvidia GeForce RTX 2070 GPU
under $B=3, M=64, P=2, L=8$, $s_l=3$ for $l\in\{1, 2,3,\cdots, 8\}$, $n_l=64$ for $l\in \{1, 2,3,\cdots, 7\}$, $n_0=n_8=2$
and the time costs are $2.5 \times 10^{-5}$ (LS method), $1.6 \times 10^{-4}$ (MMSE method), and $1.2 \times 10^{-4}$ (CRLD method).
Therefore, although the proposed CRLD has a higher complexity compared with the MMSE and LS methods, the associated time cost can be greatly reduced by exploiting the parallel computing of GPU.

\section{Conclusion}
This letter modeled the channel estimation as a denoising problem and developed a DReL approach for channel estimation in AmBC systems. We first designed a communication protocol and then proposed a novel CRLD-based estimation scheme, which consists of an offline training phase and an online estimation phase. The proposed CRLD adopts multiple 3D denoising blocks to intelligently exploit the spatial and temporal correlations of the pilot signals, which further improves the estimation accuracy.
Theoretical analysis was also provided to characterize the properties of CRLD.
Simulation results showed that the proposed method is able to achieve a close-to-optimal performance obtained by the MMSE method.

\bibliographystyle{ieeetr}
\setlength{\baselineskip}{10pt}

\bibliography{ReferenceSCI2}

\end{document}